\documentclass[amssymb,amsmath,preprint,aps,pra,nofootinbib]{revtex4-1}
\usepackage[colorlinks]{hyperref}
\usepackage{graphicx}
\usepackage{gensymb}
\usepackage{slashed}
\usepackage{fancyhdr}
\usepackage{feynmp-auto}
\usepackage{physics}
\usepackage{bbm}
\usepackage[T1]{fontenc}

\begin{document}
\title{Superkick Effect in Vortex Particle Scattering}

\author{Shiyu Liu$^{1,2}$}
\author{Bei Liu$^{3}$}
\author{Igor P. Ivanov$^{3}$}
\email{ivanov@mail.sysu.edu.cn}
\author{Liangliang Ji$^{1}$}
\email{jill@siom.ac.cn}
\affiliation{$^{1}$State Key Laboratory of High Field Laser Physics, Shanghai Institute of Optics and Fine Mechanics (SIOM), Chinese Academy of Sciences, Shanghai 201800, China\\
$^{2}$Center of Materials Science and Optoelectronics Engineering, University of Chinese Academy of
Sciences, Beijing 100049, People’s Republic of China\\
$^{3}$School of Physics and Astronomy, Sun Yat-sen University, Zhuhai 519082, China}
\date{\today}

\begin{abstract}
	Vortex states of photons or electrons are a novel and promising experimental tool across atomic, nuclear, and particle physics.
	Various experimental schemes to generate high-energy vortex particles have been proposed.
	However, diagnosing the characteristics of vortex states at high energies remains a significant challenge,
	as traditional low-energy detection schemes become impractical for high-energy vortex particles due to their extremely short de Broglie wavelength.
	We recently proposed a novel experimental detection scheme based on a mechanism called ``superkick''
	that is free from many drawbacks of the traditional methods and can reveal the vortex phase characteristics.
    In this paper, we present a complete theoretical framework for calculating the superkick effect in elastic electron scattering and systematically 
    investigate the impact of various factors on its visibility.
    In particular, we argue that the vortex phase can be identified either by 
    detecting the two scattered electrons in coincidence or by analyzing the characteristic azimuthal asymmetry in individual final particles.
\end{abstract}

\maketitle

\section{Introduction}
\label{sec: intro}
Since the 1990s, it has been repeatedly demonstrated that particles propagating in free space 
can possess an intrinsic orbital angular momentum (OAM) along the propagation direction
\cite{allen1992orbital,bliokh2007semiclassical,
uchida2010generation,verbeeck2010production,lee2019laguerre,sarenac2022experimental,luski2021vortex,bliokh2017theory,ivanov2022promises}.
These states are known as vortex particles.
Vortex particles are characterized by a wave function with a phase factor 
$\psi(\vb{r})\propto  e^{i\ell\phi_r}$, 
where $\phi_r$ denotes the azimuthal angle in the transverse plane
\cite{bliokh2017theory,ivanov2022promises}.
As the azimuthal angle $\phi_r$ is not defined on the central axis of the transverse plane, 
this axis represents the phase singularity axis of the vortex state, where the amplitude of the wave function vanishes.
For scalar particles, the vortex state is an eigenstate of the orbital angular momentum operator 
$\hat{L}_z=-i\hbar\frac{\partial}{\partial \phi_r}$, with eigenvalue $\ell \hbar$; 
for particles with spin, 
the vortex state is an eigenstate of both the total angular momentum and the helicity\cite{ivanov2022promises}.

Scattering of vortex particles has been extensively studied in the last decade; see, for example, the reviews
\cite{bliokh2017theory,ivanov2022promises}.
Since vortex particles carry an intrinsic OAM, they provide a novel degree of freedom for investigating atomic physics
\cite{afanasev2013off,scholz2014absorption}, 
nuclear physics \cite{zadernovsky2006excitation,wu2022dynamical}, and particle physics
through scattering \cite{afanasev2022delta,jentschura2011generation}. 
To make use of vortex particles in nuclear and particle physics, 
their energy must reach the scale of MeV to GeV.
Although experimental production of high-energy vortex particles has not yet been achieved, 
physicists have proposed several practical approaches to address this challenge.
In 2011, U. D. Jentschura and V. G. Serbo proposed a method to 
produce high-energy twisted photons by Compton backscattering of twisted laser photons off ultrarelativistic electrons
\cite{jentschura2011generation,jentschura2011compton}. 
Subsequently, generating high-energy vortex photons via 
nonlinear Thomson scattering \cite{taira2017gamma,chen2019generation} and nonlinear Compton scattering \cite{PhysRevD.109.016005} 
have also been introduced.
Further immersing a cathode in a solenoid field provides 
an efficient and versatile approach to generating high-energy electron vortex states\cite{floettmann2020quantum,bu2024generation}. 
This approach is expected to enable the output of vortex electron beams at approximately 200 MeV\cite{10624179,zhuhai2024}.

The precise characterization of vortex properties presents another significant challenge 
once high-energy vortex particles are successfully generated in experiments.
For low-energy particles, one commonly employed technique is the fork grating diffraction method
\cite{saitoh2013measuring,guzzinati2014measuring}.
Nevertheless, in the case of high-energy particles, their extremely short de Broglie wavelengths render 
the diffraction method impractical for detecting vortex phase.

Recently a scheme for diagnosing vortex 
particles in the high-energy regime is developed \cite{li2024unambiguous}, by leveraging the superkick effect\cite{barnett2013superweak,afanasev2022superkicks,PhysRevA.105.013522,liu2023threshold}.
In this scheme, vortex particles elastically scatter with a tightly focused non-vortex probe particle beam.
The presence of the vortex phase can be inferred from the anomalous momentum shift of the final-state particles induced by the superkick effect.
It is a universal kinematic phenomenon that is independent of the type of scattered particle and can thus be extended to other vortex particles.

In this paper, we take M{\o}ller scattering (elastic electron-electron scattering) as an example to provide 
an in-depth and systematic discussion on the superkick effect, 
including the theoretical and numerical methods, the dependence on various key parameters, realistic experimental considerations and so on. This paper is organized as follow: In Sec. \ref{superkick}, we briefly introduce the superkick effect and the detection scheme.
Sec. \ref{Theoretical} presents the theoretical and numerical methods we adopt, 
including exact numerical calculations and analytical expressions derived under the high-energy paraxial approximation.
In Sec. \ref{Impact}, we discuss the influence of various factors on the superkick effect, 
focusing primarily on wave packet size, alignment jitter, non-collinear beam collisions, spin-flip effects etc.
Finally, Sec. \ref{conclusion}, a conclusion will be given.

\section{The Superkick Effect}
\label{superkick}

\begin{figure}[h]
    \centering
    \includegraphics[height=0.4\textwidth]{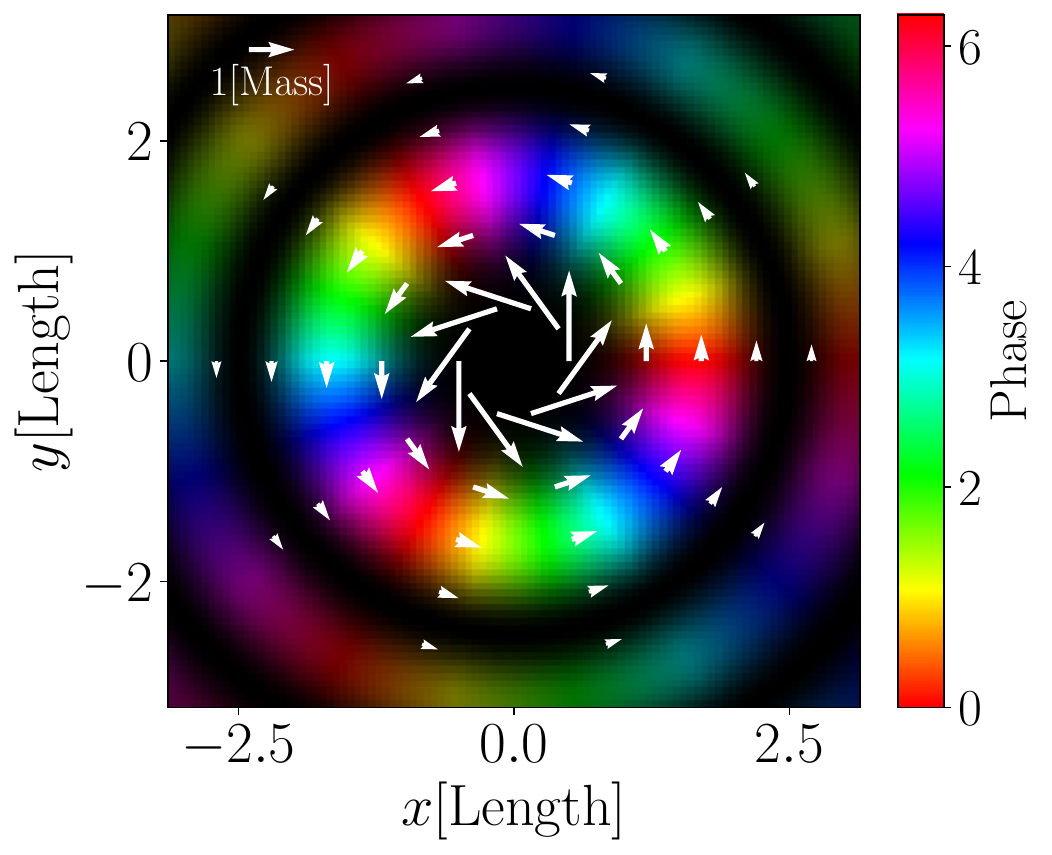}
    \caption{The transverse distribution of the wavefunction at $z=0$ and the distribution of azimuthal momentum flow in vortex particles.
    The white arrows indicate the local momentum in the azimuthal direction $\vb{p}_\phi(\vb{r})$. 
    In the figure, the color hue represents the phase, while the brightness denotes the relative probability density.}
    \label{local}
\end{figure}
\begin{figure}[h]
    \centering
    \includegraphics[height=0.3\textwidth]{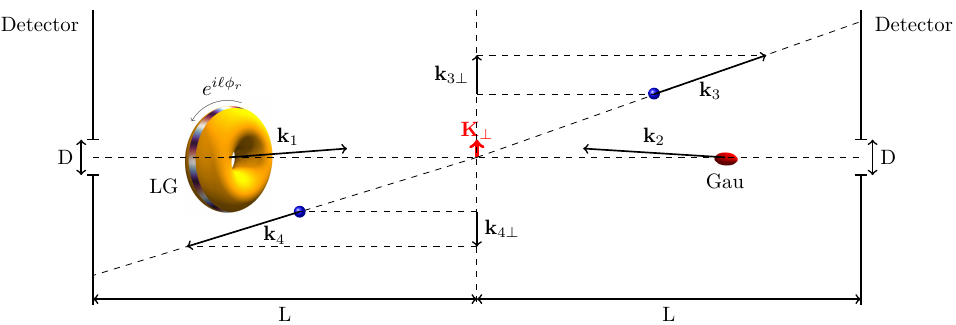}
    \caption{Schematic diagram of the experimental detection. Here, LG denotes Laguerre-Gaussian
    vortex electrons, and Gau represents tightly focused Gaussian wave-packet electrons.
    Detectors are placed at a distance $L$ on either side of the collision point to detect final-state scattered particles.
    Each detector has a small hole $D$ at its center to allow the passage of incident particles. 
    $\vb{k}_1$ and $\vb{k}_2$ denote the typical plane wave components of the LG and Gaussian wave packets, respectively.
    $\vb{k}_3$ and $\vb{k}_4$ denote the scattered plane wave electrons.}
    \label{Fig6}
\end{figure}
In the wave function of a vortex particle, the vortex phase $e^{i\ell\phi_r}$ 
induces a localized component of transverse momentum, 
$\vb{p}_\phi(\vb{r})=\vb{j}_{\phi}(\vb{r})/\abs{\psi(\vb{r})}^2=\vb{e}_\phi\,\ell/r$, in the azimuthal direction\cite{berry2013five}.
In Fig. \ref{local}, we illustrate a scalar particle in a Bessel vortex state as an example, 
showing the distribution of its probability density and local transverse momentum flow.
From a semiclassical perspective, positioning a point-like test particle in the vicinity of the vortex axis induces a transverse momentum transfer exceeding the typical value carried by the vortex particle.
This surprisingly large momentum transfer was dubbed the "superkick"\cite{barnett2013superweak}. 
This situation leads to a paradox: 
the momentum absorbed by the point-like probe particle can be much greater than the 
actual momentum of the vortex particle, resulting in a violation of momentum conservation\cite{barnett2013superweak}. 
To resolve this paradox, Ref. \cite{barnett2013superweak,PhysRevA.105.013522} suggested that both the vortex particle 
and the probe particle should be described within the quantum framework. Hence, all initial states are represented
as wave packets and the total energy-momentum conservation law is imposed as shown in Eq. \eqref{S-matrix}. 
Each plane-wave component of the wave packet obeys the law of momentum conservation.

This highly localized momentum of a vortex particle can be probed via elastic scattering with normal particles of compact wave packets. 
As shown in \cite{li2024unambiguous}, elastic scattering between vortex and ordinary electrons is proposed to reveal this effect. 
A detailed setup is illustrated  in  Fig. \ref{Fig6}. Here a tightly focused Gaussian electron wave packet collides with a vortex electron with a non-zero impact parameter. 
The superkick effect manifests as follows:
For an initially tightly focused Gaussian electron wave packet, the mean transverse momentum $\langle k_{y} \rangle$ is zero. 
However the localized transverse momentum $K_y$ of the vortex electron can induce an offset in the center of distribution of the superposed momentum  for both scattered particles $k_{3y}+k_{4y}=K_y$.
Furthermore, as dictated by the direction of the local transverse momentum, the offset direction is oriented perpendicular to the impact parameter $\vb{b}$. 
The momentum distribution of final electrons can be observed by detectors arranged off-axis relative to the collinear orientation, 
allowing the overall momentum shift to be identified, thereby revealing the vortex phase structure.

\section{Theoretical and Numerical Methods}
\label{Theoretical}
In M{\o}ller scattering, when the initial-state electrons are not plane waves but possess a wave packet distribution, the S-matrix element is given by:
\begin{align}
    S_{fi}=\int\frac{\dd^3\vb{k}_1}{(2\pi)^32E_{k_1}}\frac{\dd^3\vb{k}_2}{(2\pi)^32E_{k_2}}
    \phi_1(\vb{k}_1)\phi_2(\vb{k}_2)(2\pi)^4\delta^4(k_1+k_2-K)\mathcal{M}(\vb{k}_1,\vb{k}_2;\vb{k}_3,\vb{k}_4)\label{S-matrix},
\end{align}
where $\vb{k}_1$ and $\vb{k}_2$ denote the momenta of the incident electrons, while $\vb{k}_3$ and $\vb{k}_4$ denote those of the scattered electrons. 
The total four-momentum of the final-state particles is given by $K^\mu=k_3^\mu+k_4^\mu$, with the total energy $E_K=E_{k_3}+E_{k_4}$ and total momentum $\vb{K}=\vb{k}_3+\vb{k}_4$.
$\phi_1(\vb{k}_1)$ and $\phi_2(\vb{k}_2)$ represent the momentum distributions of the incident particle wave packets and are normalized as  
\begin{align}
    \int\frac{\dd^3\vb{k}}{(2\pi)^32E_{\vb{k}}}\abs{\phi_i(\vb{k})}^2=1.
\end{align}

In Eq. \eqref{S-matrix}, $\mathcal{M}$ denotes the invariant matrix element associated with the process under consideration. In our calculation, 
this corresponds to the invariant matrix element for Møller scattering, whose explicit form is given in Refs. \cite{greiner2008quantum,berestetskii1982quantum}.

\subsection{Numerical Method}
In numerical calculations, explicit accounting for the specific forms of the wave packet $\phi_i(\vb{k})$ and the invariant matrix element $\mathcal{M}$ is not required.  
By exploiting the properties of the delta function, the six-fold integral in Eq. \eqref{S-matrix} can be reduced to a two-fold integral, yielding the following expression:  
\begin{align}
    S_{fi}=\int \frac{\dd\Omega_{k_1}}{16\pi^2}
    \frac{\vb{k}_1^2}{\abs{E_K k_1-E_{k_1}K\cos\theta}}\phi(\vb{k}_1)\phi(\vb{k}_2)\mathcal{M}(\vb{k}_1,\vb{k}_2;\vb{k}_3,\vb{k}_4),\label{simplify}
\end{align}
In Eq. \eqref{simplify}, the energy $E_{k_1}$ is given by  
\begin{align}
    E_{k_1}=\frac{sE_K+\sqrt{K^2\cos^2\theta[s^2+4m^2(E_K^2-K^2\cos^2\theta)]}}{2(E_K^2-K^2\cos^2\theta)}, 
\end{align}
where $s=K^\mu K_\mu=E_K^2-K^2$ and $\theta$ denoting the angle between $\vb{K}$ and $\vb{k}_1$.  
For Eq. \eqref{simplify}, numerical methods such as the Newton-Cotes formulas or the trapezoidal rule can be used for computation \cite{scherer2017computational}.  

The calculation of the differential probability requires six-dimensional phase-space integration  
\begin{align}
    \dd W=\abs{S_{fi}}^2\frac{\dd^3\vb{k}_3}{(2\pi)^3 2E_{k_3}}\frac{\dd^3\vb{k}_4}{(2\pi)^3 2E_{k_4}}.
\end{align}
The Monte Carlo integration method \cite{lepage2021adaptive} was employed to efficiently compute the differential scattering probability.

\subsection{Analytical Model-Relativistic Regime and Paraxial Approximation}
We employ the Laguerre-Gaussian (LG) wave packet and the 
Gaussian wave packet to describe initial-state electrons $\phi_1$ and $\phi_2$, respectively
\begin{align}
    \phi_1(\vb{k}_1)\propto\frac{(\sigma_{1\perp}k_{1\perp})^\ell}{\sqrt{\ell !}}
    \exp[-\frac{k_{1\perp}^2\sigma_{1\perp}^2}{2}-\frac{(k_{1z}-p_{1z})^2\sigma^2_{1z}}{2}+i\ell\phi_{k_1}],
\end{align}
\begin{align}
    \phi_1(\vb{k}_2)\propto
    \exp[-\frac{k_{2\perp}^2\sigma_{2\perp}^2}{2}-\frac{(k_{2z}-p_{2z})^2\sigma^2_{2z}}{2}+i\vb{b}_\perp\cdot \vb{k}_{2\perp}].
\end{align}
We assume that the transverse ($\sigma_{1\perp}$ and $\sigma_{2\perp}$) and longitudinal dimensions 
($\sigma_{1z}$ and $\sigma_{2z}$) of the wave packets are on the nanometer scale \cite{PhysRevD.101.076009}, 
and the corresponding momentum broadening is on the order of hundreds of electron volts.
We adopt the central momenta of the incident particles as $p_{1z}=10$MeV and $p_{2z}=-10$MeV, 
much larger than the momentum spread of the wave packets $p_{zi}\sigma_{zi}\gg 1$. 
In this case, the incident particles are in the strongly paraxial regime, with the divergence angles of the wave packets given by 
$k_{1\perp}/k_{1z}\sim k_{2\perp}/k_{2z}\sim10^{-5}$.

For final-state scattered electrons, the cross section in M{\o}ller scattering is proportional 
to $1/\sin^4\theta$ \cite{greiner2008quantum,berestetskii1982quantum}. 
As a result, the majority of the scattering signal is concentrated in the paraxial region. 
To effectively separate the scattered electron signal from the background, 
we analyze electrons scattered at angles in the range of 1–5 mrad. This corresponds to a transverse momentum of approximately 10 keV, which exceeds the transverse momenta of both the initial LG and Gaussian wave packets, $k_{1\perp}$ and $k_{2\perp}$, even when considering their broadening.

In the Born approximation, the amplitude of M{\o}ller scattering is given by\cite{greiner2008quantum,berestetskii1982quantum}
\begin{align}
    \mathcal{M}\propto\biggl(\frac{\bar{u}_{k_3}\gamma^\mu u_{k_1}\bar{u}_{k_4}\gamma_\mu u_{k_2}}{t}
    -\frac{\bar{u}_{k_4}\gamma^\mu u_{k_1}\bar{u}_{k_3}\gamma_\mu u_{k_2}}{u}\biggr)\label{amplitude}.
\end{align}
In the paraxial scattering regime ($\theta\ll1$), the t-channel contribution, proportional to $1/(1-\cos\theta)$, significantly higher than that of the u-channel of $1/(1+\cos\theta)$. Therefore, we consider only the former.
In the paraxial and relativistic regime, 
we evaluate the electron bispinor $u_{k_i}$ in the limit $\theta_{k_i}\rightarrow 0$, where $i=1,2,3,4$, and neglect the electron mass\cite{li2024unambiguous}. 
Then the amplitude in Eq .\eqref{amplitude} scales as $\mathcal{M}\propto 1/t$. The Mandelstam variable 
$t$ can be simplified to 
\begin{align}
    t=(k_3-k_1)^2\approx-(\vb{k}_{1\perp}-\vb{k}_{3\perp})^2.
\end{align}
With $k_{3\perp}\gg k_{1\perp}$,
the term $1/(\vb{k}_{1\perp}-\vb{k}_{3\perp})^2$ is expanded:
\begin{align}
    \frac{1}{(\vb{k}_{1\perp}-\vb{k}_{3\perp})^2}\approx
    \frac{1}{k^2_{3\perp}}\biggl[1+2\frac{k_{1\perp}}{k_3{\perp}}\cos(\phi_{k_3}-\phi_{k_1})\biggr]\label{expand}.
\end{align}
The invariant amplitude can be approximated accordingly
\begin{align}
    \mathcal{M} \propto \frac{1}{k^2_{3\perp}}\biggl[1+2\frac{k_{1\perp}}{k_3{\perp}}\cos(\phi_{k_3}-\phi_{k_1})\biggr]
    \delta_{\lambda_1\lambda_3}\delta_{\lambda_2\lambda_4}\label{ApproxAmp}.
\end{align}
The Kronecker $\delta$ in Eq. \eqref{ApproxAmp} indicates that helicities are always conserved along each fermion line.
As stated in Ref. \cite{liu2023threshold}, the differential probability is thus derived
\begin{align}
    \dd W\propto\abs{\mathcal{I}_\perp}^2\dd^2 \vb{k}_{\perp3}\dd^2\vb{K}_{\perp}\label{KdiffPro},
\end{align}
where
\begin{align}
    \mathcal{I}_{\perp}\propto\,&i^\ell \sigma_{1\perp}^\ell 
    e^{-\frac{K_{f\perp}^2\sigma_{2\perp}^2}{2}}e^{i\ell\phi_B}\frac{1}{k_{3\perp}^2}\notag\\
    &\biggl(N_{\ell+1,\ell}+\frac{1}{k_{3\perp}}i^{-1}e^{-i\phi_B}e^{i\phi_{k_3}}N_{\ell+2,\ell-1}+
    \frac{1}{k_{3\perp}}ie^{i\phi_B}e^{-i\phi_{k_3}}N_{\ell+2,\ell+1}\biggr)\label{It}.
\end{align}
In Eq. \eqref{It}, $N_{m,n}$ is defined by the Whittaker function \cite{gradshteyn2014table}
\begin{align}
    N_{m,n}(\alpha,\beta)\equiv\frac{\Gamma(m/2+n/2+1/2)}{\beta\alpha^{m/2}\Gamma(n+1)}
    \exp(-\frac{\beta^2}{8\alpha})M_{m/2,n/2}\biggl(\frac{\beta^2}{4\alpha}\biggr)
\end{align}
and 
\begin{align}
    \phi_B=\arctan(\frac{b_y-iK_{y}\sigma_{2\perp}^2}{b_x-iK_{x}\sigma_{2\perp}^2}).
\end{align}

\subsection{Benchmark}
The results from numerical methods and analytical approximations are benchmarked considering a total angular momentum of 11/2 for the vortex-state electron. 
The incident LG wave packet has a transverse width of $\sigma_{1\perp}$=10nm and a longitudinal width of $\sigma_{1z}$=5nm,
while the Gaussian wave packet has $\sigma_{2\perp}$=2nm and $\sigma_{2z}$=1nm.
\begin{figure}[h]
    \centering
    \includegraphics[height=0.4\textwidth]{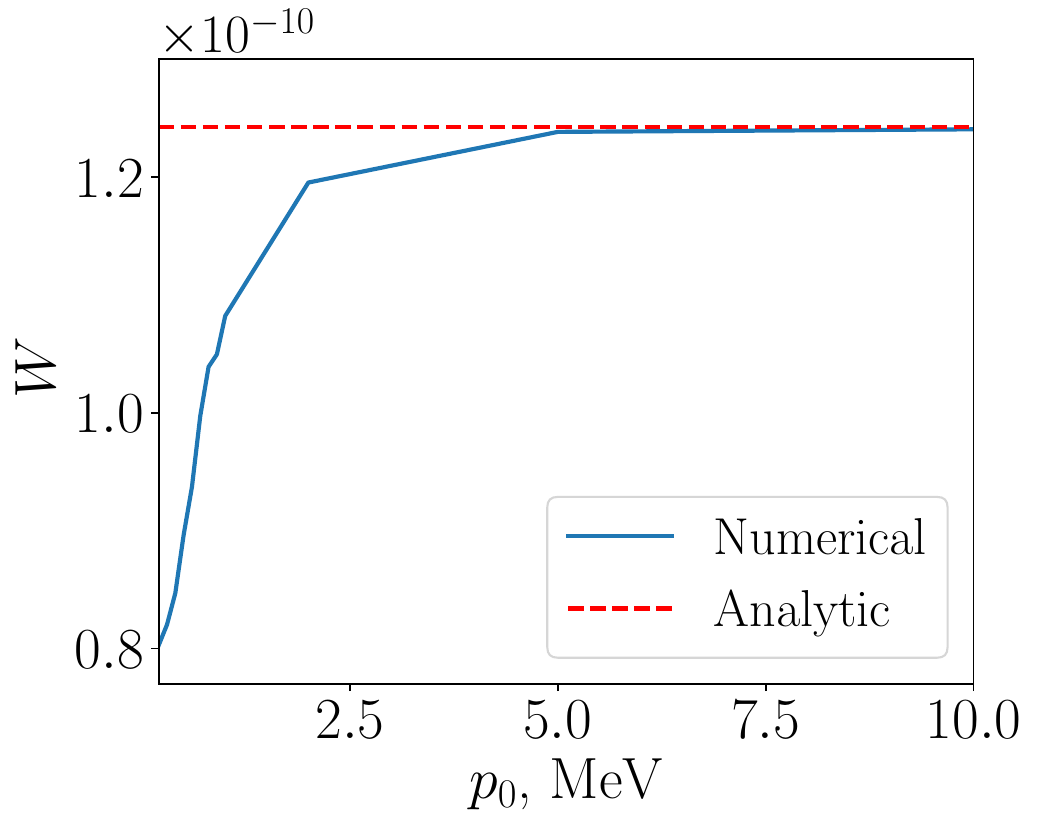}
    \caption{The total probability $W$ as a function of the incident electron momentum $p_0$. 
    The impact parameters are set to $b_x=20$nm and $b_y=0$nm.
    The term 'Numerical' refers to results obtained from the numerical method, 
    while 'Analytic' denotes results derived under the relativistic paraxial approximation. 
    The integration range of $k_{3\perp}$ is fixed from 10 keV to 50 keV.}
    \label{bench}
\end{figure}
Fig. \eqref{bench} compares the probability from both methods at different incident electron momentum  $p_0$.  
When the electron momentum is significantly larger than its rest mass ($p_0>5$MeV), the analytical results match the numerical one perfectly, 
as clearly demonstrated by Eqs. \eqref{KdiffPro} and \eqref{It}.
However, in the weakly relativistic regime, while the analytical value remains constant, 
the numerical one declines significantly, suggesting that the approximation is no longer applicable.
Thus, when considering experimental validation using weakly relativistic electrons, such as those from a transmission electron microscope (TEM), 
we rely on the numerical method for evaluation.
Here although the overall probability decreases the superkick effect remains observable.

\section{Impact of Various Factors}
\label{Impact}
\begin{figure}[h]
    \centering
    \includegraphics[height=0.3\textwidth]{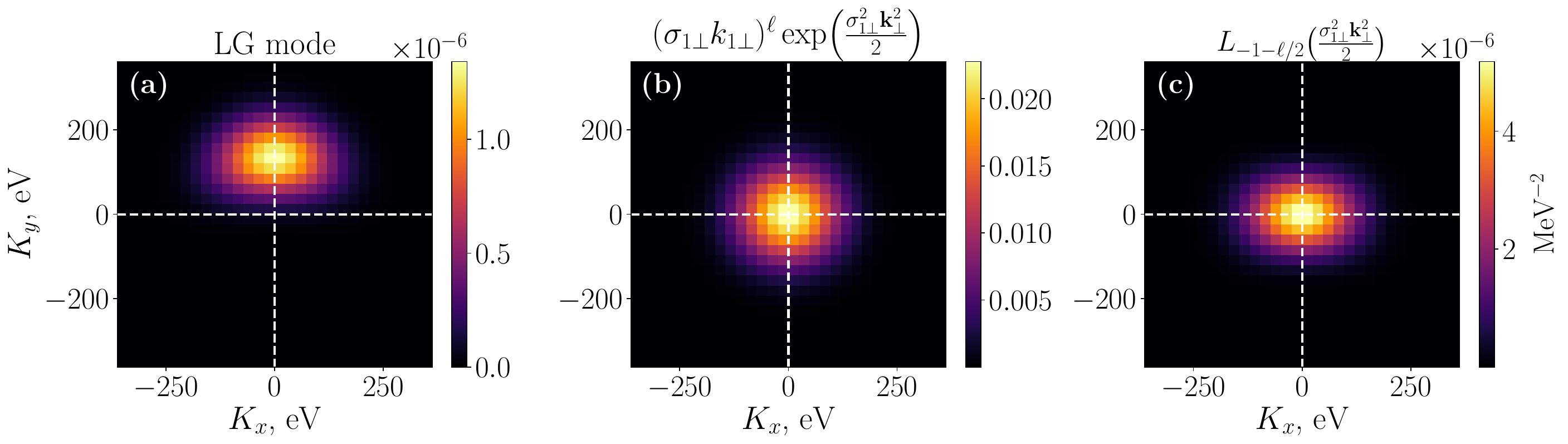}
    \caption{The total transverse momentum distribution is shown, with the collision parameter
    $\vb{b}$ fixed at $b_x=5$nm and $b_y=0$nm for panels (a) through (c). The incident electron energy is 10 MeV.
    The transverse momentum distribution of the wave packet is specified in the title of each subfigure.}
    \label{Fig9}
\end{figure}

The superkick effect manifests itself in the differential scattering probability as a function of the total transverse momentum $\vb{K}$ of the final-state particles
\begin{align}
    w(\vb{K}_{\perp})=\frac{\dd W}{\dd^2\vb{K}_{\perp}}=
    \int \abs{S_{fi}}^2\frac{\dd^3 \vb{k}_3\dd k_{4z}}{(2\pi)^64E_{k_3}E_{k_4}}\label{Wk}.
\end{align}
Fig. \ref{Fig9} presents the detailed results using the parameters from Fig. \ref{bench}. The total transverse momentum exhibits a noticeable displacement
perpendicular to $b_x$. This shift vanishes if one removes the vortex phase in the momentum space of the LG wave packet (Fig. \ref{Fig9}(b) ) 
or in the coordinate space  (\ref{Fig9}(c)) . 

\begin{figure}[h]
    \centering
    \includegraphics[height=0.37\textwidth]{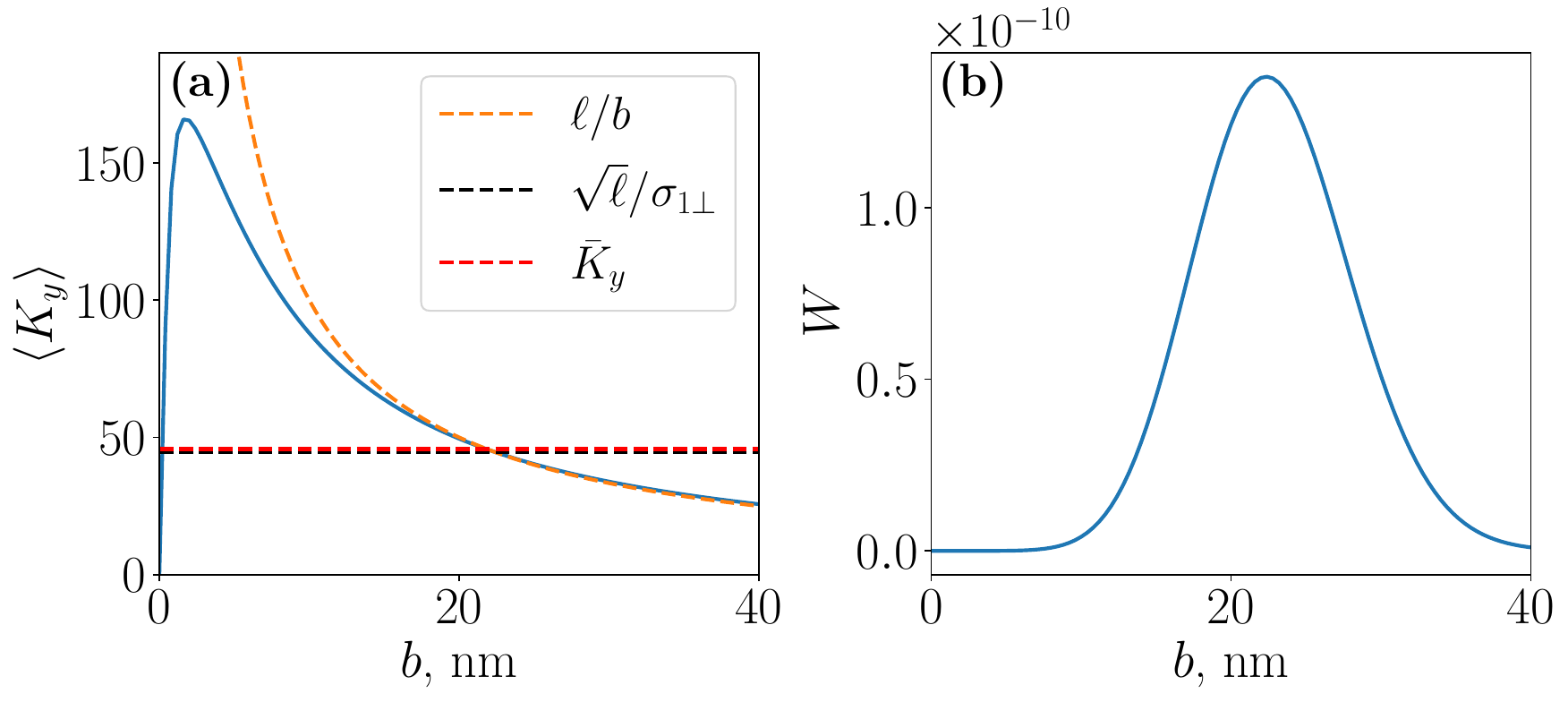}
    \caption{(a): Average transverse momentum 
    $\langle K_y \rangle$ as a function of the impact parameter $b_x$.
    The black dashed line, $\sqrt{\ell}/\sigma_{1\perp}$, represents the characteristic transverse momentum of the LG vortex electron.
    The orange dashed line corresponds to the local transverse momentum, $\frac{\ell}{b} \vb{e}_y$, 
    induced by the vortex phase. (b): Scattering probability as a function of the impact parameter $b_x$.}
    \label{KyExpect}
\end{figure}
Fig. \ref{KyExpect}(a) shows the expectation value of the transverse momentum $\langle K_y \rangle$ of the scattered particles as a function of the impact parameter $b_x$. 
For small values of $b_x$, the average transverse momentum $\langle K_y \rangle$
goes beyond the characteristic transverse momentum of the vortex particles, $\sqrt{\ell}/\sigma_{1\perp}$.
In semiclassical picture, the momentum diverge at center. However, in the quantum picture, it converges to 0 at $b_x=0$. This effect arises due to the wave packet nature of the probe particle.
When a finite-sized wave packet approaches the phase singularity, 
it encounters symmetric transverse momenta from various azimuthal directions, which cancel each other out,
thus preventing the divergence of $\ell/b$.  
As the impact parameter $b_x$ increases, $\langle K_y \rangle$ gradually approaches $\ell/b$, 
in agreement with the momentum transfer predicted by the semiclassical picture. 

In Fig. \ref{KyExpect}(b), we present the scattering probability for different impact parameters. One may ask how momentum is conserved when superkick happens. This phenomenon arises due to the localized tangential momentum of vortex phase structure. Although $\langle K_y \rangle$ exceeds the transverse momentum $\sqrt{\ell}/\sigma_{1\perp}$ for small impact radii, it gradually drops below this value as $b$ increases. In other words,  it is a redistribution of momentum induced by the localized kick. By weighting $\langle K_y \rangle$ with the scattering probability $W$ and integrating over $b$, we obtain the average transverse momentum $\bar{K}_y=\int \langle K_y \rangle W\dd b/\int W\dd b$. 
This corresponds to the red dashed line in Fig. \ref{KyExpect}(a). We observe that $\bar{K}_y$ exactly matches the characteristic transverse momentum of the vortex particle, i.e., momentum is conserved averagely in this process.

The comparison in Fig. \ref{Fig9} is straightforward. 
However, in more realistic scenarios, it is essential to account for various factors that have non-negligible effects on the results, 
including the alignment jitter of electron sources, collisions at non-collinear geometries, the finite size of wave packets and other related considerations. 
We therefore conduct a systematic analysis of these factors to provide guidance for future experiments. 
Unless explicitly stated otherwise, the main parameters follow those in Fig. \ref{Fig9}.

\subsection{Effects of Alignment Jitter and Statistics}
In realistic experiments, it is impossible to precisely control the impact parameter of two colliding electron beams, as alignment jitter is always present, especially when signals are collected from an adequate number of scattering events.

\begin{figure}[h]
    \centering
    \includegraphics[height=0.32\textwidth]{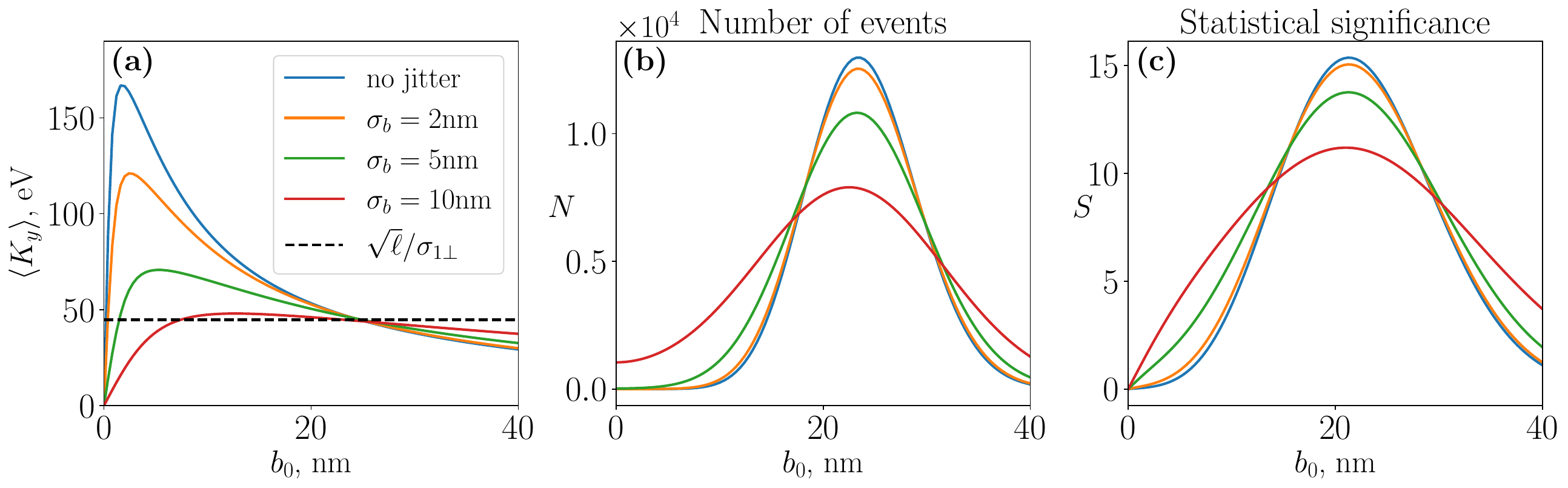}
    \caption{(a): The average transverse momentum $\langle K_y \rangle$  is shown as a function of the impact parameter
     $b_0$ for different fluctuation amplitudes $\sigma_b$.
    (b): The number of scattered particles varies as a function of the impact parameter  $b_0$.
    (c): The statistical significance of a non-zero $\langle K_y \rangle$ 
    is analyzed as a function of the impact parameter $b_0$.
    The transverse size of the Gaussian wave packet is $\sigma_{2\perp}=2$nm. 
    The collision parameters are defined as $b_x=b_0$ and $b_y=0$.}
    \label{Fig7}
\end{figure}
The jitter is modeled using the following distribution:
\begin{align}
    f(\vb{b};\vb{b}_0)=
    \frac{1}{\pi\sigma_{b}^2}\exp[-\frac{(\vb{b}-\vb{b}_0)^2}{\sigma_{b}^2}],
\end{align}
where $\vb{b}_0$  is the center of the distribution and 
$\sigma_{b}$ represents its characteristic width.
After accounting for the distribution, 
the smeared differential probability $\bar{w}(\vb{K}_{\perp})$ is defined as:
\begin{align}
    \bar{w}(\vb{K}_{\perp};\vb{b}_0)=
    \int w(\vb{K}_{\perp};\vb{b})f(\vb{b};\vb{b}_0)\dd^2\vb{b}.
\end{align}
The distribution of the 
averaged transverse momentum $\langle K_y \rangle$ 
as a function of  $b_0$ of the jitter radii $\sigma_{b}$ is shown in Fig.\ref{Fig7}(a). Apparently, the superkick effect gradually diminishes as the jitter becomes more significant.

In this scheme, the interactions between electrons within each electron beam should be neglected.
Therefore, it is required that spacing among the electrons is larger than the size of electron wave packets.
For instance, with an average current of 16 nA electron source, their average spacing  is about 3 mm, sufficiently large as compared to the wave packet size. 
An experiment running for $10^3$
seconds yields $N_0 = 10^{14}$ electron collision attempts.
As shown in Fig. \ref{Fig7}(b), 
approximately $10^4$ scattered electron signals around $b\approx\sigma_{1\perp}$ are expected.
The detector response ($5\times 10^6$ electrons per pixel per second) \cite{dijkstra2002overview,dubey2024high} is sufficient for our purpose.

As illustrated in Fig. \ref{KyExpect}, the total transverse momentum shows a 
deviation of 150 eV. This naturally raises the question of how well such a small momentum shift can be resolved in an electron beam of 10 MeV.
Importantly, we do not directly measure the tiny momentum shift. Instead, the transverse momenta of the scattered particles, $\vb{k}_{3\perp}$ and $\vb{k}_{4\perp}$, whose magnitudes are on the order of 10 keV, are recorded. This is well within the range of realistic experimental detectors.
By simultaneously measuring both $\vb{k}_{3\perp}$ and $\vb{k}_{4\perp}$, we can determine the total transverse momentum $\vb{K}_\perp=\vb{k}_{3\perp}+\vb{k}_{4\perp}$.

In addition, when detecting final-state particles, their momentum cannot be precisely determined. 
Therefore, consideration of the transverse momentum uncertainty is essential. 
Here, we set $L$ and $D$ to 2m and 4mm, respectively, as shown in the experimental schematic in Fig. \ref{Fig6}.
In the experimental proposal, we consider using a pixelated silicon (Si) electron detector \cite{dijkstra2002overview,dubey2024high}, 
which offers resolution of 70 $\mu$m with each pixel.
This corresponds to a transverse momentum uncertainty of 
$\delta k_{\perp}=350$ eV for detecting the scattered electrons.

Taking into account alignment jitter, signal statistics, and the finite momentum resolution in the detection of final-state particles, 
we define the statistical significance $S$
\begin{align}
    S=
    \frac{\langle K_y \rangle\sqrt{N(b_0)}}{\sqrt{\langle K_y^2 \rangle+(\delta k_{\perp})^2}},
\end{align}
to evaluate the outcome.
The effect is deemed experimentally observable when the statistical significance $S$ is greater than 5.
As shown in Fig. \ref{Fig7}(c),  although the superkick effect is strong at smaller values of  $b$, 
the limited number of scattering signals poses a challenge for experimental detection, leading to a lower statistical significance.
Taking into account all relevant factors, the optimal impact position is $b\sim \sqrt{\ell}\sigma_{1\perp}$, which maximizes statistical significance.
Furthermore, we find that even if the position uncertainty $\sigma_b$ is 
comparable to the transverse size of the LG wave packet $\sigma_{1\perp}$, 
a nonzero $\langle K_y \rangle$ can still be experimentally observed.

\subsection{Effects of Non-collinear Collision}
\begin{figure}[h]
    \centering
    \includegraphics[height=0.2\textwidth]{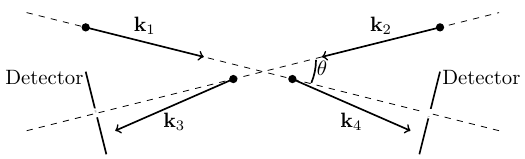}
    \caption{Schematic of a non-collinear collision setup, where the crossing angle $\theta$ is defined}
    \label{CrossingAngle1}
\end{figure}
\begin{figure}[h]
    \centering
    \includegraphics[height=0.35\textwidth]{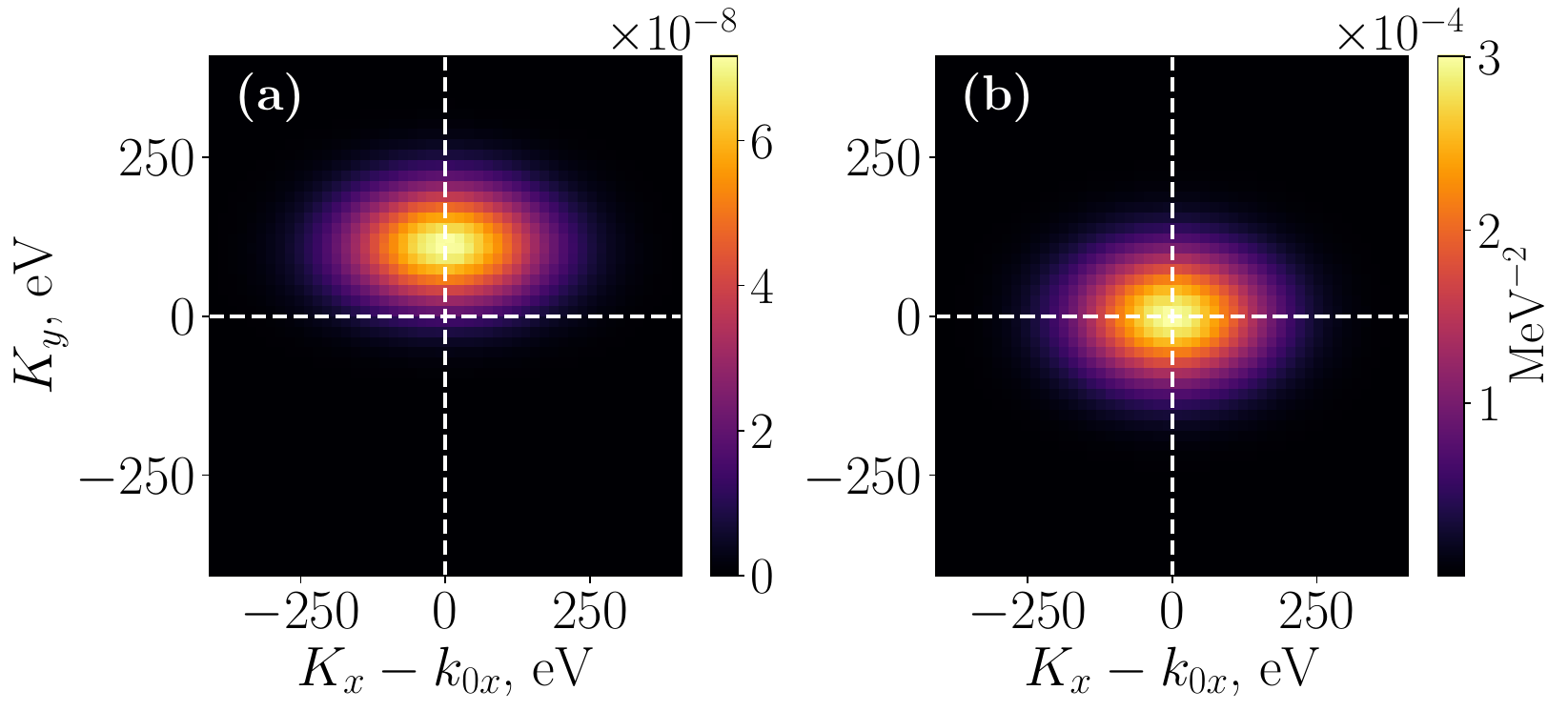}
    \caption{(a):Distribution of $w(\vb{K}_{\perp})$ at a crossing angle of 45 degrees. (b): 
    Distribution of $w(\vb{K}_{\perp})$ at the collision of two Gaussian wave packets for the same crossing angle. 
    In both cases, the collision parameters are set as $b_x=5$nm and $b_y=0$nm. The horizontal axis has been offset by 
    subtracting the momentum change in the x-direction due to a cross collision, given by $k_{0x}=k_0\sin(3\pi/4)\approx 7.07$ MeV.}
    \label{Fig11}
\end{figure}

As shown in Fig. \ref{CrossingAngle1}, in a more realistic experimental setup, 
in order to separate the two beamlines, the collisions can be arranged at an appropriate nonzero crossing angle.
The distribution of the differential scattering probability $w(\vb{K}_{\perp})$ for a crossing angle of $\pi/4$ has been obtained, as illustrated in Fig. \eqref{Fig11}(a). 
The center of the total transverse momentum $K_x$ shifts from 0 to $k_{0x}=k_0\sin(3\pi/4)\approx 7.07$ MeV 
due to the cross collision. In contrast, the shift along the $K_y$-direction is evident due to the superkick effect.
For comparison, we analyzed the collision between two Gaussian wave packets at the same crossing angle, as shown in Fig. \eqref{Fig11}(b).
In this case, the displacement of the distribution center along the $K_y$-direction vanishes.

\begin{figure}[h]
    \centering
    \includegraphics[width=0.7\linewidth]{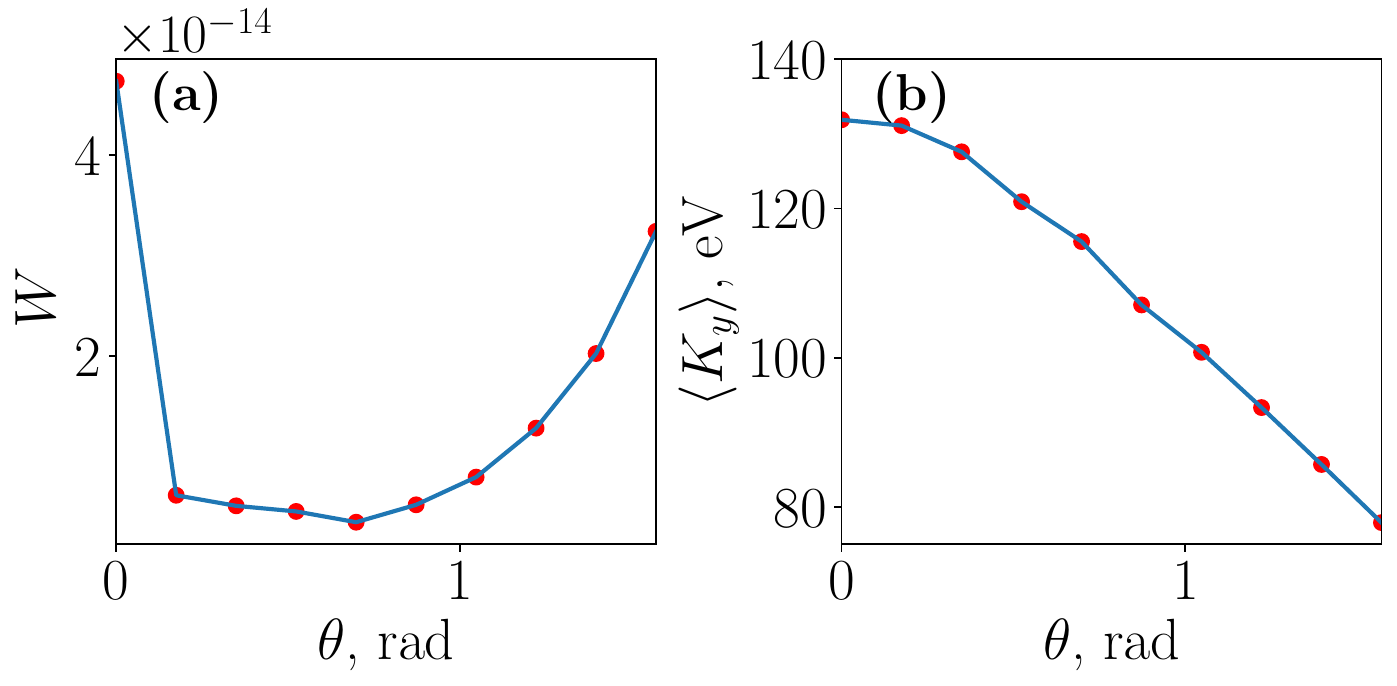}
    \caption{The scattering probability (a) and the averaged total transverse momentum $\langle K_y\rangle$ (b) are analyzed as a function of the crossing angle under non-collinear collisions. We set the impact parameters to $b_x=5$nm and $b_y=0$nm. }
    \label{ObliqueProPy}
\end{figure}

We further examine the effect of the crossing angle $\theta$, and the results are presented in Fig. \ref{ObliqueProPy}.
Fig. \ref{ObliqueProPy}(a) shows that as the crossing angle increases, the scattering probability declines and then gradually rises up. 
This is because in non-collinear collisions the wave packets overlap for a shorter duration at a small crossing angles. 
When the crossing angle further increases, the tightly focused Gaussian wave packet passes through both sides of the 'ring' of the LG vortex particle, where its probability density peaks. 
However, the superkick effect, denoted by $\langle K_y\rangle$, decreases as the angle increases. In non-collinear collisions,  the trajectory of the probe crosses a line related to varying impact parameters, thus the local transverse momentum shall be averaged, which leads to lower values at larger angles.

\subsection{Effects of the Wave Packet Size of the Gaussian Probe}
\begin{figure}[h]
    \centering
    \includegraphics[height=0.30\textwidth]{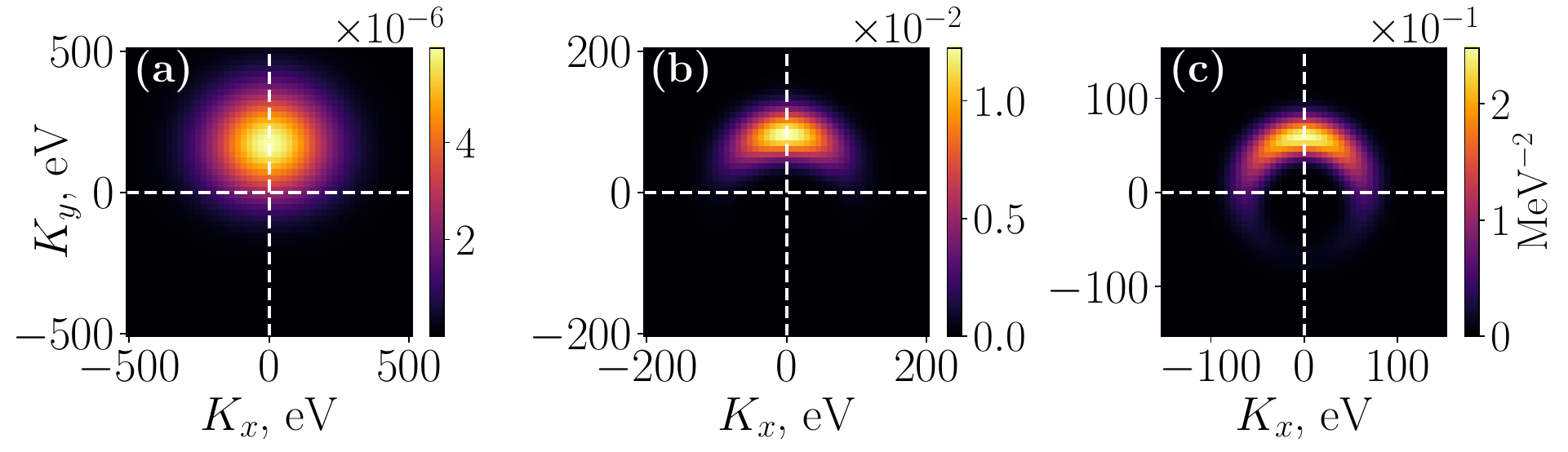}
    \caption{(a)-(c):The Gaussian wave packets have transverse sizes of 1, 5, and 10 nm respectively.
    The impact parameters are $b_x=$5 nm and $b_y=$0 nm.}
    \label{Fig4}
\end{figure}
\begin{figure}[h]
    \centering
    \includegraphics[height=0.45\textwidth]{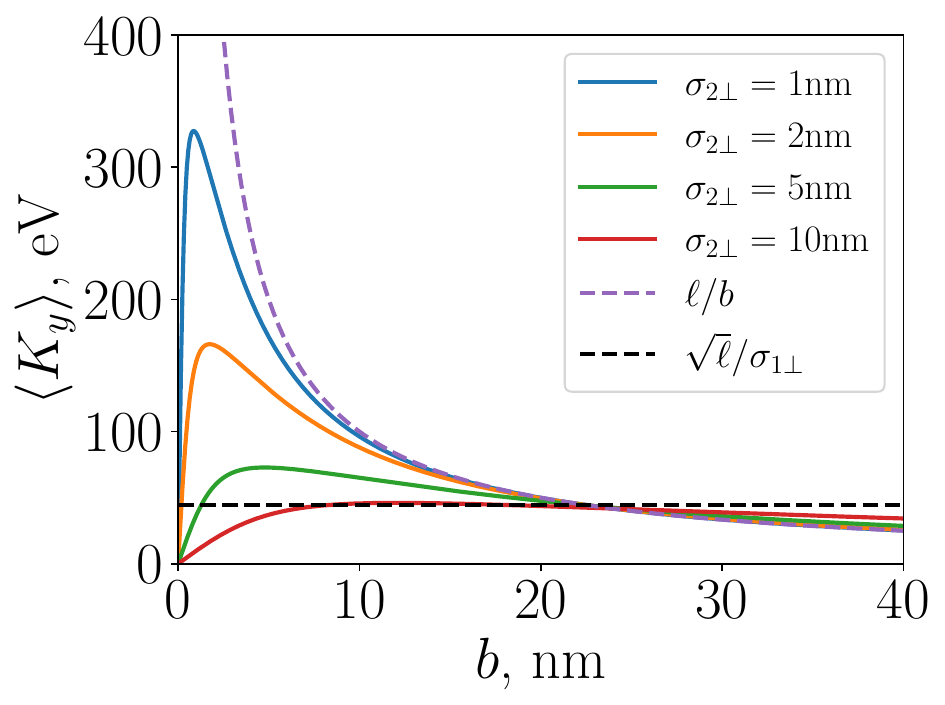}
    \caption{Effect of different transverse sizes of Gaussian wave packets on the average transverse momentum.}
    \label{Fig5}
\end{figure}
In general, the superkick effect is more significant when the probe is more localized.
Fig. \ref{Fig4} presents the differential scattering probability by changing transverse sizes of Gaussian wave packets, following
\begin{align}
    w(\vb{K}_{\perp})=\frac{\dd W}{\dd^2\vb{K}_{\perp}}=
    \int \abs{S_{fi}}^2\frac{\dd^3 \vb{k}_3\dd k_{4z}}{(2\pi)^64E_{k_3}E_{k_4}}\label{Wk}.
\end{align}
The momentum displacements are centered around 200, 100, and 50 eV in \ref{Fig4}(a), \ref{Fig4}(b), 
and \ref{Fig4}(c), for wave packet sizes of 1, 5 and 10 nm, respectively. 
As the Gaussian wave packet size is approaching to the LG one, the distribution exhibits annular shape, a reflection of the vortex intensity profile.

The above trend is clearly seen in the $b_x$-dependent 
distribution in Fig. \ref{Fig5}. When the transverse size of the Gaussian wave packet is set to
$\sigma_{2\perp}=\sigma_{1\perp}=10$nm, 
the total mean transverse momentum is slightly below the characteristic value $\sqrt{\ell}/\sigma_{1\perp}$ 
of the LG vortex; hence, the superkick effect disappears.

\subsection{The Azimuth Distribution of the Scattered Particles}
In addition to the total transverse momentum distribution of final-state particles, we further consider the angular distribution of one of the scattered particles, $k_3$. 
This dependence can also be found in Eq. \eqref{ApproxAmp} and Eq.\eqref{It} . 
The distribution is defined as follows:
\begin{align}
    \Phi(\phi_{k_3})=\frac{\dd W}{\dd\phi_{k_3}}=\int\abs{S_{fi}}^2\frac{k_{3\perp}\dd k_{3\perp}\dd k_{3z}\dd^3\vb{k}_4}{(2\pi)^64E_{k_3}E_{k_4}}.
\end{align}
\begin{figure}[h]
    \centering
    \includegraphics[height=0.35\textwidth]{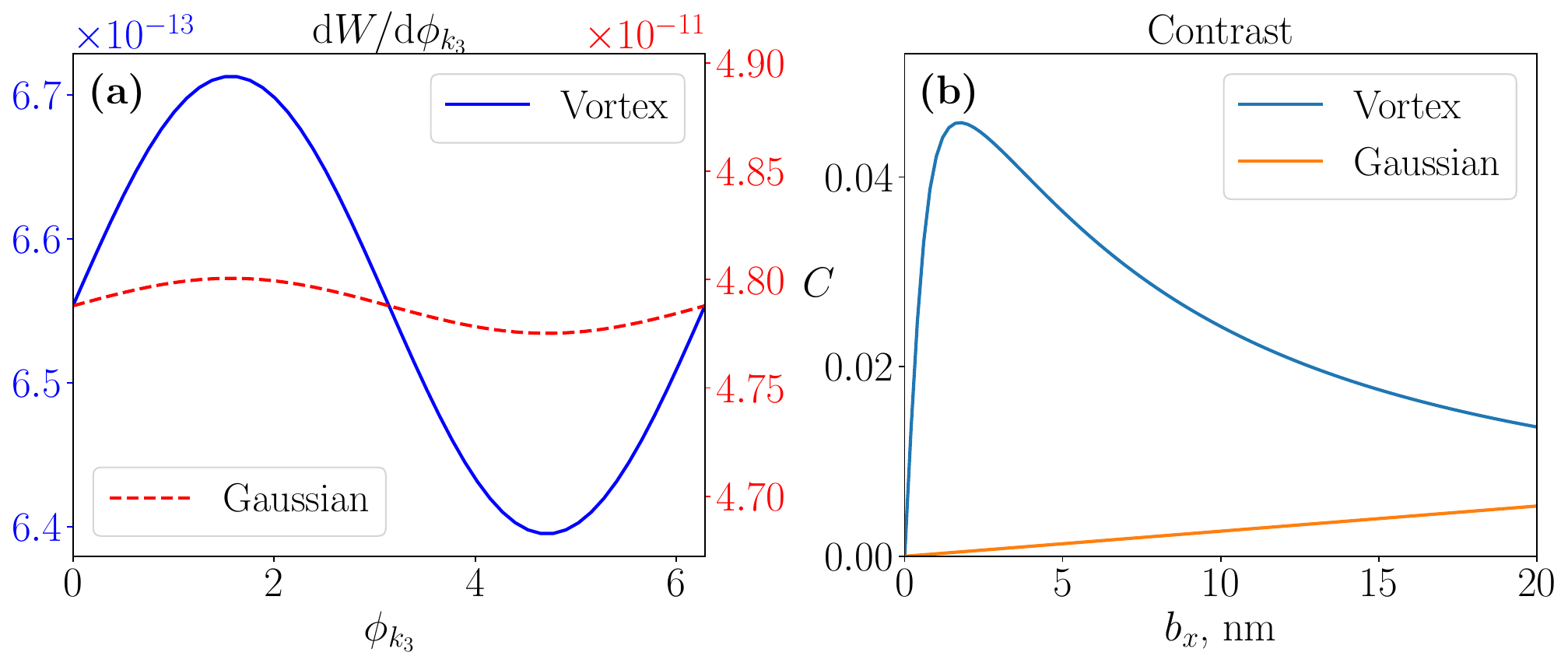}
    \caption{(a): The distribution of the differential scattering probability with respect to the azimuthal angle $\phi_{k_3}$ of the scattered electron.
    the impact parameters are set to $b_x=10$nm and $b_y=0$nm. 
    Vortex particle scattering is represented by the left vertical axis, while Gaussian wave packet scattering is represented by the right vertical axis.
    (b): Dependence of the contrast C on the impact parameter $b_x$.}
    \label{Phi3}
\end{figure}
Fig. \ref{Phi3}(a) shows that for a nonzero impact parameter, the azimuthal distribution of the final-state electron follows a sinusoidal pattern. It appears for both Gaussian wave packet 
collisions and vortex scattering but more pronounced in the latter. 
To characterize this phenomenon, we define the contrast $C$ of the distribution to calibrate the degree of  variation
\begin{align}
    C=\frac{\Phi_{\text{max}}-\Phi_{\text{min}}}{\Phi_{\text{max}}+\Phi_{\text{min}}}.
\end{align}
In Fig. \ref{Phi3}(b), while the contrast is zero at $b_x=0$nm in both cases, we find that the one in vortex scattering is greater by an order of magnitude than that in Gaussian scattering. Further, a peak appears around $b\sim 2$ nm for vortex scattering. 
The distinctive trend in azimuth distribution provides an additional approach to identify the superkick effect.

\subsection{The Spin Flip Effect}
It is also of interest to determine whether spin flip plays a significant role in this process.
Eq. \eqref{S-matrix} shows that the superkick effect originates from the vortex phase of the wave packet, 
whereas spin flip originates from the invariant amplitude  $\mathcal{M}$, 
with no direct coupling between them.
\begin{figure}[h]
    \centering
    \includegraphics[width=0.45\linewidth]{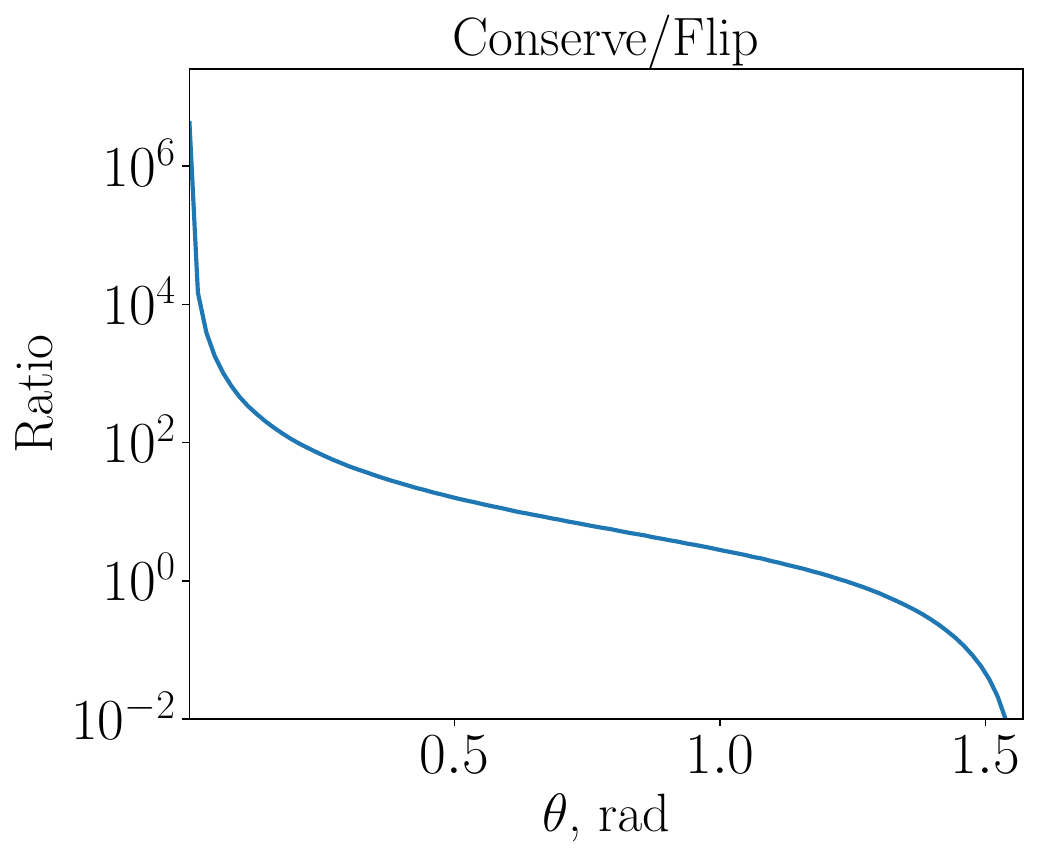}
    \caption{The ratio of the differential scattering probabilities of spin-conserving to spin-flip as a function of  $\theta_{k_3}$.
    In the figure, the azimuthal angle is fixed at $\pi/2$.}
    \label{SpinThe}
\end{figure}

Owing to the nature of M{\o}ller scattering, 
most of the scattered electrons remain in the paraxial region.
In this region, spin-flip transitions are significantly suppressed.
According to QED theory, spin conservation prevails over spin flip by a factor of  $1/\theta^2$.
In the Fig. \ref{SpinThe}, we present the ratio of the differential scattering probability 
for spin-conserving to that for spin-flip as a function of the polar angle $\theta_{k_3}$ of the scattered electron..
As shown in Fig. \ref{SpinThe}, 
in the paraxial region, the spin-conserving channel dominates over the spin-flip channel by six orders of magnitudes.
Beyond the paraxial region, the spin-flip effect becomes more obvious.
For large scattering angles  ($\theta_{k_3}\geq1.2$), the spin-conserving channel is significantly suppressed, 
allowing the spin-flip channel to dominate.

\section{Conclusions and Perspectives}
\label{conclusion}
Vortex particle scattering offers a unique degree of freedom for advancing our understanding of atomic, nuclear, and particle physics.
The detection of high-energy vortex particles is a crucial step in such studies.
To tackle this problem, we propose an experimental scheme based on the superkick effect 
that enables the characterization of the vortex properties of vortex particles \cite{li2024unambiguous}.

In this work, we present a detailed analysis of the proposed scheme. 
We provided both exact numerical calculations and analytical expressions under the relativistic paraxial approximation and investigated several factors that could potentially affect the detection efficiency, 
among which the transverse size of the wave packets and alignment jitter plays important roles.
Effects such as spin flip and the energy of the incident electron have a minor influence on the superkick effect. We also pointed out that the superkick effect can still be observed in more realistic situations such as non-collinear collision configuration.

Furthermore, we demonstrated that the vortex phase also manifests as an asymmetry in the azimuthal distribution of a single scattered electron.  Employing a single-particle detection method relaxes the stringent requirement on detector response time.

While our calculations in this paper are illustrated using Møller scattering,
the superkick effect is a ubiquitous feature of various vortex scattering processes, including Compton scattering and electron-proton scattering. Extending this idea to the detection of high-energy vortex states of photons, ions, and hadrons is a natural progression, a topic that will be explored in more detail in future work.

\section{Acknowledgments}
S.L. thanks Yujian Lin for his assistance with Fig. \ref{Fig6}. S.L. and L.J. thank Sun Yat-sen University for hospitality during the workshop 
'Vortex states in nuclear and particle physics'. S.L. and L.J. acknowledge the support by National Science Foundation of China (Grant No. 12388102), 
CAS Project for Young Scientists in Basic Research (Grant No. YSBR060), and the National Key R{\&}D Program of China (Grant No. 2022YFE0204800). 
B.L. and I.P.I. thank the Shanghai Institute of Optics and Fine Mechanics for hospitality during their visit. 

\clearpage
\bibliography{biblib}

\end{document}